\documentclass[12pt,a4paper]{article}
\usepackage{graphicx,hyperref,wrapfig}
\hypersetup{
     colorlinks=true,       
     linkcolor=red,          
     citecolor=green,        
     filecolor=magenta,      
     urlcolor=blue           
}
\newcommand{\newc}{\newcommand}
\def\u#1{\verb!#1!\endgroup}

\newc{\HW}{\textsf{HERWIG}}
\newc{\TAUOLA}{\textsf{TAUOLA}}
\newc{\ThePEG}{\textsf{ThePEG}}
\newc{\HWPP}{\textsf{Herwig++}}
\newc{\evt}{\textsf{EvtGen}}
\newc{\fortran}{\textsf{FORTRAN}}
\newc{\decayer}{\textsf{Decayer}}

\newc{\HWPPClass}[1]{\href{http://projects.hepforge.org/herwig/doxygen/classHerwig_1_1#1.html}{\textsf{#1}}}
\newc{\ThePEGClass}[1]{\href{http://projects.hepforge.org/thepeg/doxygen/classThePEG_1_1#1.html}{\textsf{#1}}}
\newc{\HWPPParameter}[2]{\href{http://projects.hepforge.org/herwig/doxygen/#1Interfaces.html\##2}{{\bf #2}}}
\newc{\ThePEGParameter}[2]{\href{http://projects.hepforge.org/thepeg/doxygen/#1Interfaces.html\##2}{{\bf #2}}}
\newc{\HWPPParameterValue}[3]{\href{http://projects.hepforge.org/herwig/doxygen/#1Interfaces.html\##2}{{\bf [#2=#3]}}}
\newc{\ThePEGParameterValue}[3]{\href{http://projects.hepforge.org/thepeg/doxygen/#1Interfaces.html\##2}{{\bf [#2=#3]}}}

\begin{document}
\tolerance=100000
\thispagestyle{empty}
\setcounter{page}{0}
 \begin{flushright}
CERN-PH-TH/2008-213\\
CP3-08-51\\
IPPP/08/75\\
DCPT/08/150\\
KA-TP-27-2008\\
December 2008
\end{flushright}
\begin{center}
{\Large \bf Herwig++ 2.3 Release Note}\\[0.7cm]

M.~B\"ahr$^1$,
S.~Gieseke$^1$,
M.~Gigg$^2$,
D.~Grellscheid$^2$,
K.~Hamilton$^3$,
S.~Pl\"atzer$^1$,
P.~Richardson$^{2}$,
M.~H.~Seymour$^{4,5}$,
J.~Tully$^2$,

E-mail: {\tt herwig@projects.hepforge.org}\\[1cm]

$^1$\it Institut f\"ur Theoretische Physik, Universit\"at Karlsruhe.\\[0.4mm]
$^2$\it IPPP, Department of Physics, Durham University. \\[0.4mm]
$^3$\it Centre for Particle Physics and Phenomenology, Universit\'e Catholique de Louvain.\\[0.4mm] 
$^4$\it Physics Department, CERN.\\[0.4mm]
$^5$\it School of Physics and Astronomy, University of Manchester.\\[0.4mm]
\end{center}

\vspace*{\fill}

\begin{abstract}{\small\noindent A new release of the Monte Carlo
    program \HWPP\ (version 2.3) is now available. This version includes
    a number of improvements including: the extension of the program to
    lepton-hadron collisions; the inclusion of several processes
    accurate at next-to-leading order in the POsitive Weight
    Hardest Emission Generator~(POWHEG) scheme; the inclusion of
    three-body decays and finite-width effects in Beyond the Standard
    Model~(BSM) physics processes; a new procedure for reconstructing
    the kinematics of the parton shower based on the colour structure of
    the hard scattering process; a new model for baryon decays including
    excited baryon multiplets; the addition of a soft component to the
    multiple scattering model of the underlying event; new matrix
    elements for Deep Inelastic Scattering~(DIS) and $e^+e^-$ processes.
  }
\end{abstract}

\tableofcontents
\setcounter{page}{1}

\section{Introduction}

The last major public version (2.2) of \HWPP, is described in great detail
in \cite{Bahr:2008pv,Bahr:2008tx}. This release note therefore only lists the
changes which have been
made since the last release~(2.2). The manual has been updated to 
reflect these changes and this release note is only intended to highlight these
new features and the other minor changes made since the last version.

Please refer to \cite{Bahr:2008pv} and the present paper if
using version 2.3 of the program.

The main new features of this version are 
the extension of the program to lepton-hadron collisions, the inclusion of
several processes accurate at next-to-leading order~(NLO) in the POWHEG scheme,
the extension of our simulation of BSM physics processes to include finite-width
effects and three-body decays,
a new procedure for reconstructing the kinematics of the parton shower
based on the colour structure of the hard scattering process,
a new model for baryon decays including excited baryon multiplets,
the addition of a soft component to the multiple scattering model of
the underlying event, the inclusion of the matrix elements for
charged and neutral current processes in DIS and Higgs boson production via
gauge boson fusion processes in lepton-lepton collisions.

Some code which we use to test and develop \HWPP\ but we do not
expect to be required by the vast majority of users has been moved from
the core parts of the program and included in the new {\tt Contrib} directory
together with some code we had not previously released. It is intended that in
future this directory will both provide a repository of examples and allow us to 
distribute external modules.
In addition a number of other changes have been
made and a number of bugs have been fixed.

\subsection{Availability}

The new program, together  with other useful files and information,
can be obtained from the following web site:
\begin{quote}\tt
       \href{http://hepforge.cedar.ac.uk/herwig/}{http://hepforge.cedar.ac.uk/herwig/}
\end{quote}
  In order to improve our response to user queries, all problems and requests for
  user support should be reported via the bug tracker on our wiki. Requests for an
  account to submit tickets and modify the wiki should be sent to 
  {\tt herwig@projects.hepforge.org}.

  \HWPP\ is released under the GNU General Public License (GPL) version 2 and 
  the MCnet guidelines for the distribution and usage of event generator software
  in an academic setting, which are distributed together with the source, and can also
  be obtained from
\begin{quote}\tt
 \href{http://www.montecarlonet.org/index.php?p=Publications/Guidelines}{http://www.montecarlonet.org/index.php?p=Publications/Guidelines}
\end{quote}

\section{Lepton-Hadron Collisions and Kinematic Reconstruction}

  Rather than the simple procedure used in previous versions of the program which 
  preserved the centre-of-mass energy and rapidity of the final-state system
  in initial-state
  parton showers and the momentum of the system for final-state showers we now
  use a procedure that preserves the properties of colour-singlet systems where possible
  as originally intended in Ref.~\cite{Gieseke:2003rz}. For example in Higgs production
  via vector boson fusion we would now preserve the momenta of the two virtual
  gauge bosons. The new procedure \HWPPParameterValue{QTildeReconstructor}{ReconstructionOption}{Colour}
  is now the default, however for many processes, for example Drell-Yan, it is identical to the
  previous procedure. The procedure used is described in
  more detail in~\cite{UpdatedManual}.

  As a consequence of this extension the kinematic reconstruction of DIS events is now possible for the first time, an example of using \HWPP\ for DIS is provided with the release.

\section{Powheg}

  A number of processes are now included at next-to-leading order in the POWHEG scheme 
  of Refs.~\cite{Nason:2004rx,Frixione:2007vw} which allows the generation of events with
  NLO accuracy while only generating positive weight events.
  Our implementation includes both the generation
  of the kinematics of the hard process scattering process with NLO accuracy, the generation
  of the hardest emission and the full treatment of soft wide angle radiation, more details
  can be found in Refs.~\cite{UpdatedManual,Hamilton:2008pd}.
  The Born-level process is generated with NLO accuracy by the 
\begin{itemize}
\item \HWPPClass{MEqq2gZ2ffPowheg} class for the production and decay of the $\gamma^*/Z^0$
      boson in the Drell-Yan process;
\item \HWPPClass{MEqq2W2ffPowheg} class for the production and decay of the $W^\pm$
      boson in the Drell-Yan process;
\item \HWPPClass{MEPP2HiggsPowheg} class for the production of the Higgs boson via the 
      gluon-gluon fusion process;
\item \HWPPClass{MEPP2WHPowheg} class for the production of the Higgs boson in association 
      with the $W^\pm$ boson;
\item \HWPPClass{MEPP2ZHPowheg} class for the production of the Higgs boson in association 
      with the $Z^0$ boson.
\end{itemize}
  The parton shower in these processes is then generated using the \HWPPClass{PowhegEvolver}
  which inherits from the standard \HWPP\ \HWPPClass{Evolver} and implements both the generation
  of the truncated shower and the hardest emission in the event using the:
\begin{itemize}
\item \HWPPClass{DrellYanHardGenerator} class for processes with an intermediate vector boson;
\item \HWPPClass{GGtoHHardGenerator} class for Higgs boson production via gluon-gluon fusion.
\end{itemize}
  Examples of using these new processes can be found in the {\tt LHC-Powheg.in}
  and {\tt TVT-Powheg.in} example files supplied with the release for the LHC and 
  Tevatron.

\section{BSM Physics}

  The previous version of \HWPP\ included the automatic generation of both $2\to2$
  scattering processes and $1\to2$ decays in BSM physics models. This has now been
  extended~\cite{Gigg:2008yc} to include the automatic generation of
  three-body decays. In addition whereas previously all the BSM particles were
  produced on-shell we have extended the mechanism used to generate
  off-shell hadron decays to BSM processes~\cite{Gigg:2008yc}.
  These new features are described in more detail in 
  Refs.~\cite{UpdatedManual,Gigg:2008yc}.

  In addition we have made a number of changes to \ThePEG\ which introduces a
  new layer of abstraction in the hierarchy of the vertex classes. This makes
  it much easier to add new vertices which do not have the standard perturbative
  Lorentz structure. The mechanisms for handling BSM models can now handle any
  Lorentz structure for the vertices.

\section{New Matrix Elements}

  A number of new matrix elements are included in this release:
\begin{itemize}
\item the \HWPPClass{MENeutralCurrentDIS} and \HWPPClass{MEChargedCurrentDIS} classes for the simulation
      of neutral and charged current processes in lepton-hadron collisions;
\item the \HWPPClass{MEee2HiggsVBF} class for the simulation
      of $e^+e^-\to h^0e^+e^-$ and $e^+e^-\to h^0\nu_e\bar{\nu}_e$ via gauge boson
      fusion;
\item the \HWPPClass{MEMinBias} class for soft non-perturbative scatterings in hadron-hadron collisions
      which is solely intended for use in the improved underlying event model.
\end{itemize}

\section{Baryon Decays}

  We now include a new model for baryon decays including updated particle properties
  and decays, matrix elements for many important decays including spin correlations
  and include several excited baryon multiplets for the first time. This is
  intended to have the same sophistication for the modelling of the baryons as we
  had already for the mesons. 

  By default we have included a number of additional baryon multiplets which are
  produced in the hadronization phase and slightly improve the agreement with
  LEP data. In association with this we have retuned the default parameters to 
  improve the description of LEP and B-factory data.
  A full description can be found in \cite{UpdatedManual}.

\section{Underlying Event}

  The underlying event simulation of \HWPP\ now includes two new
  features. The first one may be summarized by the term \emph{double
  parton scattering}, which refers to events where two independent hard
  scatterings can be observed. The second new feature is the inclusion
  of non-perturbative partonic scatterings into the existing underlying
  event model~\cite{Bahr:2008dy}. Thereby we replace the existing
  transverse momentum cut-off, $p_t^{\rm min}$, by a matching scale and
  describe the entire transverse momentum spectrum. This enables \HWPP\ to
  describe minimum bias data for the first time.

  \subsection{Double parton scattering}
  
  This feature allows the user to specify a given number of hard
  scatterings that are simulated in each event in addition to the
  regular underlying event. Double/multiple-parton scattering signatures
  like several high-$p_t$ jets or $b$-quark pairs or several $W$-pairs
  can be simulated using this functionality. Every production process
  can be accompanied by a set of kinematical cuts, which are freely
  selectable and completely independent of the corresponding cuts of
  other processes that are simultaneously simulated. We describe the
  configuration needed to enable this simulation in \cite{UpdatedManual}.

  \subsection{Soft component}

  \begin{wrapfigure}{r}{0.5\columnwidth}
    \includegraphics[%
      width=0.45\textwidth]{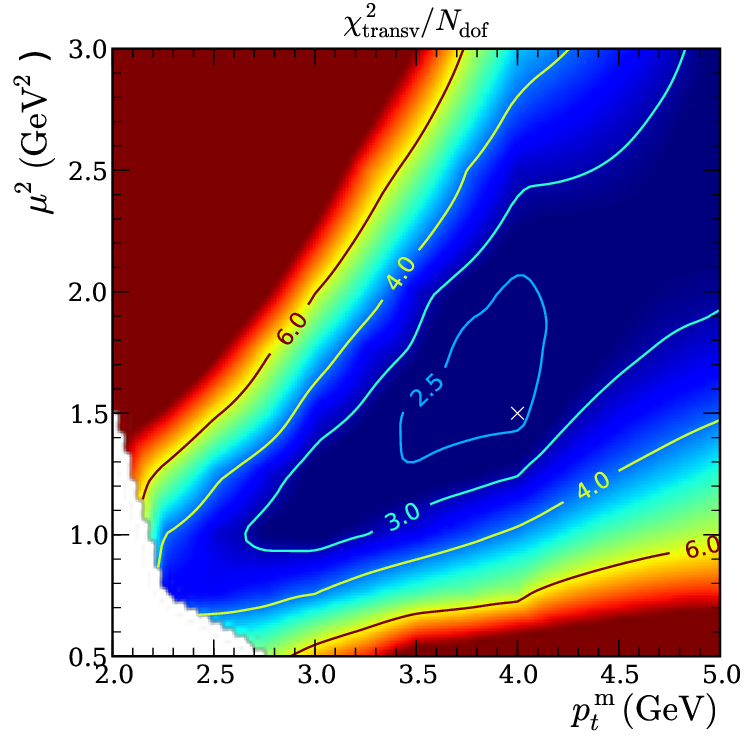}
    \caption{
      \label{fig:scans_all}
      Contour plots for the $\chi^2$ per degree of freedom of describing
      the transverse region from \cite{Affolder:2001xt}. The cross
      indicates the location of our preferred tune and the white area
      consists of parameter choices where the elastic $t$-slope and the
      total cross section cannot be reproduced simultaneously.}
  \end{wrapfigure}
\begin{figure}[!p]
    \begin{center}
      \includegraphics[%
	width=.9\columnwidth,keepaspectratio]{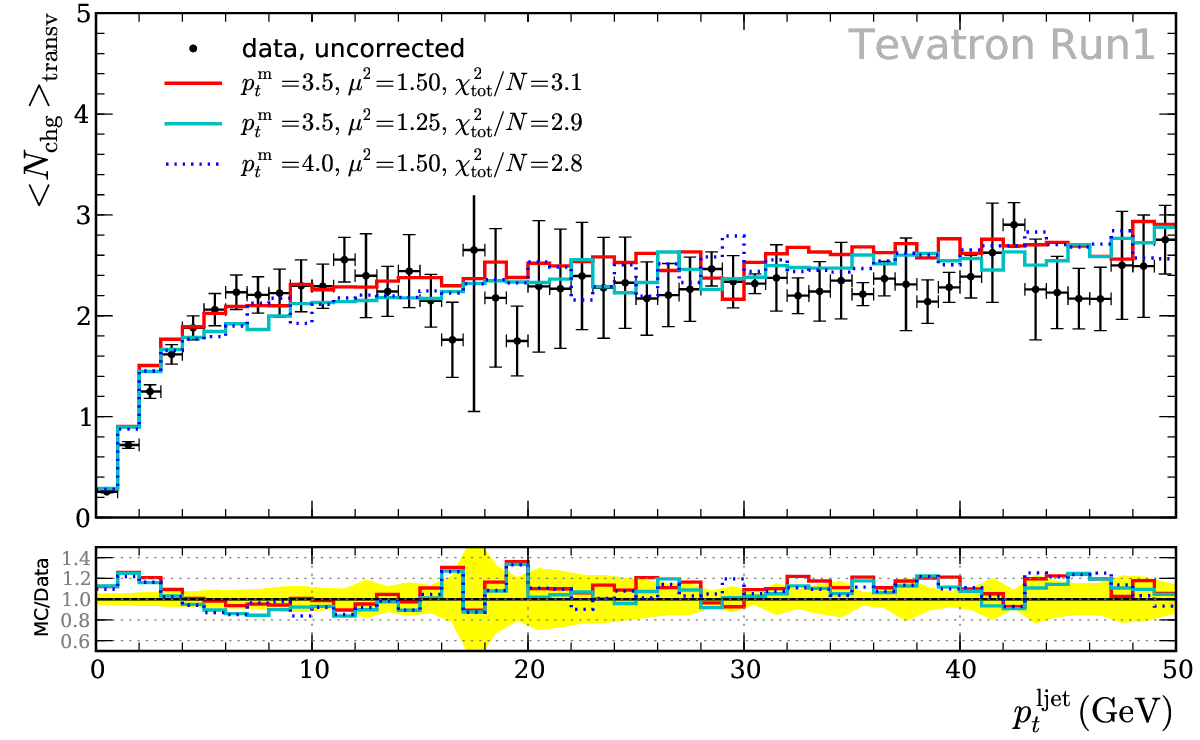}
      \\[0.5cm]
      \includegraphics[%
	width=.9\columnwidth,keepaspectratio]{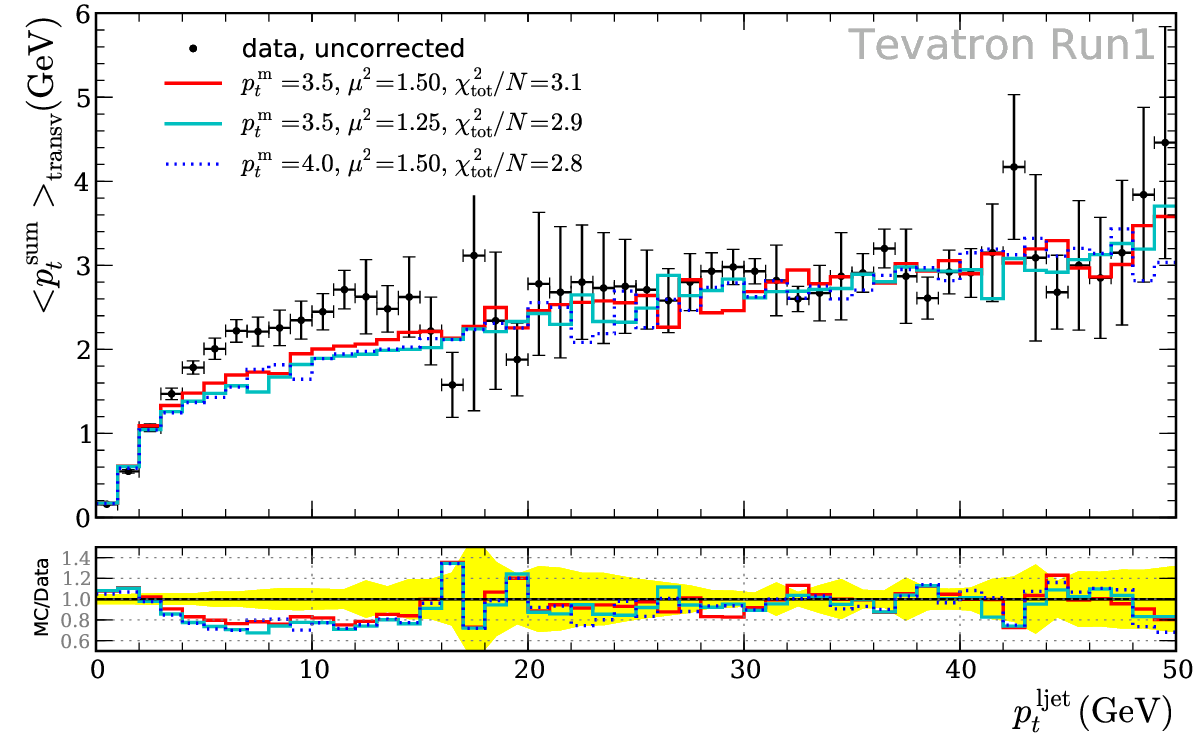}
    \end{center}
    \caption{
      \label{fig:transv_soft}
      Multiplicity and transverse momentum sum in the transverse
      region. CDF data are shown as black circles. The histograms show
      \HWPP\ with the improved model for semi-hard and soft additional
      scatters using the MRST 2001 LO \cite{Martin:2001es} PDFs for three
      different parameter sets. The lower plot shows the ratio Monte Carlo
      to data and the data error band. The legend on the upper plot shows
      the total $\chi^2$ for all observables.}
  \end{figure}

  We have implemented the inclusion of additional partonic scatters
  \emph{below} the transverse momentum cut-off along the lines of
  Ref.~\cite{Borozan:2002fk}. Additional improvements include:
  \begin{itemize}
    \item Automatic determination of the additional parameters in the
      \emph{soft} sector for any centre-of-mass energy.
    \item The possibility of different partonic overlap distributions for
      semi-hard and soft scatters. The soft overlap distribution is
      automatically fixed by the requirement to match existing data on
      the elastic $t$-slope. For energies beyond 2 TeV we use the
      parametrization of Ref.~\cite{Donnachie:1992ny} as we do for the
      total cross section. This option resolves eventual inconsistencies
      between measurements of the elastic $t$-slope and the effective
      cross section in double parton scattering events as anticipated in
      Ref.~\cite{Bahr:2008wk}.
  \end{itemize}
  The simulation has been tuned to data on the underlying event from CDF
  \cite{Affolder:2001xt}. For this comparison we could now make use of
  all available data, i.e. the jet \emph{and} minimum bias
  sample. Figure~\ref{fig:scans_all} displays the $\chi^2$ values of
  describing this data as colour code in our two-dimensional parameter
  space. A detailed comparison to data is shown in
  Fig.~\ref{fig:transv_soft}.
  
\section{Contrib}

  Starting with this release we have included a number of external modules
  which use \HWPP\ in the {\tt Contrib} directory supplied with the release.
  These will hopefully be of use to some users but are not expected 
  to be needed by most users and are not supported at the
  same level as the main \HWPP\ code. In the current release these are mainly
  modules we used to test and develop \HWPP\ and were either previously not
  released or have been removed from the main release. A full description 
  of the available modules can be found in \cite{UpdatedManual}.

\section{Other Changes}

A number of other more minor changes have been made.
The following changes have been made to improve the physics 
simulation:
\begin{itemize}
\item The \HWPPClass{IncomingPhotonEvolver} has been added to allow the simulation
      of partonic processes with incoming photons in hadron collisions.
\item A new \ThePEGClass{KTRapidityCut} class has been added to allow cuts 
      on the $p_T$ and rapidity, rather
      than just the $p_T$ and pseudorapidity used in the \ThePEGClass{SimpleKTCut}
      class. This is now used by default for cuts on massive particles
      such as the $W^\pm$, $Z^0$ and Higgs bosons and the top quark.
\item Several changes to the decayers of $B$ mesons have been made both to 
      resolve problems with the modelling of partonic decays and improve agreement
      with $\Upsilon(4s)$ data.
\item Changes have been made to allow the value of either \textsf{SCALUP} for
      Les Houches events or the scale of the hard process for internally 
      generated hard processes to be used rather
      than the transverse mass of the final-state particles 
      as the maximum transverse momentum for radiation in the parton shower.
\item The defaults for the intrinsic transverse momentum in hadron collisions
      have been reduced to $1.9$\,GeV, $2.1$\,GeV and $2.2$\,GeV for the Tevatron
      and LHC at 10\,TeV and 14\,TeV, respectively.
\item The \textsf{Pdfinfo} object is now created in the \textsf{HepMC} interface.
      However in order to support all versions of \textsf{HepMC} containing this
      feature the PDF set is not specified as not all versions contain this information.
\item The option of only decaying particles with lifetimes below a user
      specified value has been added.
\item New options for the cut-off in the shower have been added and some
      obsolete parameters removed.
\item The option of switching off certain decay modes in BSM models has been added.
\item A \ThePEGClass{Matcher} has been added for the Higgs boson to allow
      cuts to be placed on it.
\item The diffractive particles, which were not used, have been deleted from
      our default input files.
\end{itemize}
A number of technical changes have been made:
\begin{itemize}
\item Some \ThePEGClass{AnalysisHandler} classes comparing to LEP data have been
      renamed, {\it e.g.} \HWPPClass{MultiplicityCount} becomes
      \HWPPClass{LEPMultiplicityCount} to avoid confusion with
      those supplied in {\tt Contrib} for observables at the $\Upsilon(4s)$
      resonance.
\item We have reorganised the code to remove the majority of the {\tt .icc}
      files by moving the inlined functions to the headers in an effort to
      improve compile time.
\item We have restructured the decay libraries to reduce the amount of
      memory allocation and de-allocation which improves run-time performance.
\item The switch to turn off \textsf{LoopTools} has been removed because
      \textsf{LoopTools} is now used by several core modules. As \textsf{LoopTools}
      does not work on 64-bit platforms with g77 this build option is not supported.
\item We have removed support for the obsolete version of \textsf{HepMC} supplied
      with \textsf{CLHEP} and improved the support for different units options
      with \textsf{HepMC}.
\item The \textsf{EvtGen} interface has been removed until it is more stable.
\item Support for \textsf{ROOT} which was not used has been removed.
\item The CKKW infrastructure present in previous releases has been removed
      from the release.
\item The default optimisation has been increased from {\tt -O2} to {\tt -O3}.
\item The handling of the \fortran\ compiler has been improved, mainly due to 
      improvements in the \textsf{autotools}.
\item The \textsf{FixedAllocator} use to allocate memory for \ThePEGClass{Particle}
      objects in \ThePEG\ has been removed as it had no performance benefits.
\end{itemize}

The following bugs have been fixed:
\begin{itemize}
\item Problems with the mother/daughter relations in the hard process and
      diagram selection in $W^\pm$ and $Z^0$ production in association with a 
      hard jet.
\item A minor bug in the general matrix element code for fermion-vector to fermion-scalar
      where the outgoing fermion is coloured and the scalar neutral has been fixed.
\item A bug involving the selection of diagrams in some associated squark gaugino
      processes has been fixed.
\item A bug has been fixed which lead to $h^0\to\mu^+\mu^-$ being generated rather
      than $h^0\to\tau^+\tau^-$.
\item The normalisation in the \HWPPClass{Histogram} class for non unit-weight
      events has been fixed.
\item The protection against negative PDF values, which can occur when using
      NLO PDF sets, has been improved.
\item The lifetime for BSM particles is now automatically calculated at the same
      time as the width.
\item We now treat hadrons containing a top quark in the same way as hadrons containing
      BSM particles in order to support this possibility.
\item Several ambiguous uses of unsigned int have been corrected.
\item Several variables which may have been used undefined have been initialised.
\item Several memory leaks at initialisation have been fixed.
\item The configuration of \HWPP\ now aborts if no \fortran\ compiler is found as
      this is required to compile \textsf{Looptools}.
\item Several minor floating point errors that did not affect
      results have been corrected.
\end{itemize}

\section{Summary}

  \HWPP\,2.3 is the fourth version of the \HWPP\ program with a complete simulation of 
  hadron-hadron physics and contains a number of important improvements
  with respect to the previous
  version. The program has been extensively tested against
  a large number of observables from LEP, Tevatron and B factories.
  All the features needed for realistic studies for 
  hadron-hadron collisions are now present and  we look forward to 
  feedback and input from users, especially
  from the Tevatron and LHC experiments.

  Our next major milestone is the release of version 3.0 which will be at least as
  complete as \HW\ in all aspects of LHC and linear collider simulation.
  Following the release of \HWPP\,3.0 we expect that support for the 
  {\sf FORTRAN} program will cease.

\section*{Acknowledgements} 

This work was supported by Science and Technology Facilities Council,
formerly the Particle Physics and Astronomy Research Council, the
European Union Marie Curie Research Training Network MCnet under
contract MRTN-CT-2006-035606 and the Helmholtz--Alliance ``Physics at
the Terascale''. The research of K.~Hamilton was supported
by the Belgian Interuniversity Attraction Pole, PAI, P6/11.  M. B\"ahr
acknowledges support from the ``Promotionskolleg am Centrum f\"ur
Elementarteilchenphysik und Astroteilchenphysik CETA'' and
Landesgraduiertenf\"orderung Baden-W\"urttemberg.  S. Pl\"atzer
acknowledges support from the Landesgraduiertenf\"orderung
Baden-W\"urttemberg. K.~Hamilton and P.~Richardson 
would like to thank the CERN Theoretical Physics group for their
hospitality.
  
\bibliography{Herwig++}

\providecommand{\href}[2]{#2}\begingroup\raggedright\begin{thebibliography}{10}

\bibitem{Bahr:2008pv}
M.~B\mbox{\"{a}}hr {\em et.~al.}, {\it {Herwig++ Physics and Manual}},
  \href{http://xxx.lanl.gov/abs/0803.0883}{{\tt arXiv:0803.0883}}.

\bibitem{Bahr:2008tx}
M.~B\mbox{\"{a}}hr {\em et.~al.}, {\it {Herwig++ 2.2 Release Note}},
  \href{http://xxx.lanl.gov/abs/0804.3053}{{\tt arXiv:0804.3053}}.

\bibitem{Gieseke:2003rz}
S.~Gieseke, P.~Stephens, and B.~Webber, {\it {N}ew {F}ormalism for {QCD}
  {P}arton {S}howers},  {\em JHEP} {\bf 12} (2003) 045,
  [\href{http://xxx.lanl.gov/abs/hep-ph/0310083}{{\tt hep-ph/0310083}}].

\bibitem{UpdatedManual}
M.~B\mbox{\"{a}}hr {\em et.~al.} Updated version
  (\href{http://xxx.lanl.gov/abs/0803.0883v3}{{\tt 0803.0883v3}}) of
  Ref.~\cite{Bahr:2008pv}, submitted on 3/12/2008.

\bibitem{Nason:2004rx}
P.~Nason, {\it A new method for combining {NLO} {QCD} with shower {M}onte
  {C}arlo algorithms},  {\em JHEP} {\bf 11} (2004) 040,
  [\href{http://xxx.lanl.gov/abs/hep-ph/0409146}{{\tt hep-ph/0409146}}].

\bibitem{Frixione:2007vw}
S.~Frixione, P.~Nason, and C.~Oleari, {\it {Matching NLO QCD computations with
  Parton Shower simulations: the POWHEG method}},  {\em JHEP} {\bf 11} (2007)
  070, [\href{http://xxx.lanl.gov/abs/0709.2092}{{\tt arXiv:0709.2092}}].

\bibitem{Hamilton:2008pd}
K.~Hamilton, P.~Richardson, and J.~Tully, {\it {A Positive-Weight
  Next-to-Leading Order Monte Carlo Simulation of Drell-Yan Vector Boson
  Production}},  \href{http://xxx.lanl.gov/abs/0806.0290}{{\tt
  arXiv:0806.0290}}.

\bibitem{Gigg:2008yc}
M.~A. Gigg and P.~Richardson, {\it {Simulation of Finite Width Effects in
  Physics Beyond the Standard Model}},
  \href{http://xxx.lanl.gov/abs/0805.3037}{{\tt arXiv:0805.3037}}.

\bibitem{Bahr:2008dy}
M.~B\mbox{\"{a}}hr, S.~Gieseke, and M.~H. Seymour, {\it {Simulation of multiple
  partonic interactions in Herwig++}},  {\em JHEP} {\bf 07} (2008) 076,
  [\href{http://xxx.lanl.gov/abs/0803.3633}{{\tt arXiv:0803.3633}}].

\bibitem{Affolder:2001xt}
{\bf CDF} Collaboration, A.~A. Affolder {\em et.~al.}, {\it Charged jet
  evolution and the underlying event in $p\bar{p}$ collisions at 1.8 {TeV}},
  {\em Phys. Rev.} {\bf D65} (2002) 092002.

\bibitem{Borozan:2002fk}
I.~Borozan and M.~H. Seymour, {\it {An eikonal model for multiparticle
  production in hadron hadron interactions}},  {\em JHEP} {\bf 09} (2002) 015,
  [\href{http://xxx.lanl.gov/abs/hep-ph/0207283}{{\tt hep-ph/0207283}}].

\bibitem{Donnachie:1992ny}
A.~Donnachie and P.~V. Landshoff, {\it {Total cross-sections}},  {\em Phys.
  Lett.} {\bf B296} (1992) 227--232,
  [\href{http://xxx.lanl.gov/abs/hep-ph/9209205}{{\tt hep-ph/9209205}}].

\bibitem{Bahr:2008wk}
M.~B\mbox{\"{a}}hr, J.~M. Butterworth, and M.~H. Seymour, {\it {The Underlying
  Event and the Total Cross Section from Tevatron to the LHC}},
  \href{http://xxx.lanl.gov/abs/0806.2949}{{\tt arXiv:0806.2949}}.

\bibitem{Martin:2001es}
A.~D. Martin, R.~G. Roberts, W.~J. Stirling, and R.~S. Thorne, {\it {MRST2001:
  Partons and $\alpha_{S}$ from precise deep inelastic scattering and Tevatron
  jet data}},  {\em Eur. Phys. J.} {\bf C23} (2002) 73--87,
  [\href{http://xxx.lanl.gov/abs/hep-ph/0110215}{{\tt hep-ph/0110215}}].

\end{thebibliography}\endgroup
\end{document}